\documentclass[aps,pre,superscriptaddress,amssymb,showpacs,twocolumn]{revtex4-1}
\usepackage[english]{babel}
\usepackage[utf8]{inputenc}
\usepackage[T1]{fontenc}
\usepackage{graphicx}
\usepackage{amsmath}
\usepackage{hyperref}
\usepackage{makeidx}
\usepackage{comment}
\usepackage{ulem}
\usepackage{braket}

\begin{abstract}
We study transient behaviour in the dynamics of complex systems described by a set of non-linear ODE's. Destabilizing nature of transient trajectories is discussed and its connection with the eigenvalue-based linearization procedure. The complexity is realized as a random matrix drawn from a modified May-Wigner model. Based on the initial response of the system, we identify a novel stable-transient regime. We calculate exact abundances of typical and extreme transient trajectories finding both Gaussian and Tracy-Widom distributions known in extreme value statistics. We identify degrees of freedom driving transient behaviour as connected to the eigenvectors and encoded in a non-orthogonality matrix $T_0$. We accordingly extend the May-Wigner model to contain a phase with typical transient trajectories present. An exact norm of the trajectory is obtained in the vanishing $T_0$ limit where it describes a normal matrix.
\end{abstract}

\begin{document}

\title{What drives transient behaviour in complex systems?}

\author{Jacek Grela} \email{jacek.grela@lptms.u-psud.fr} 
\affiliation{LPTMS, CNRS, Univ. Paris-Sud, Universit\'{e} Paris-Saclay, 91405 Orsay, France}
\affiliation{M. Smoluchowski Institute
of Physics and Mark Kac Complex Systems Research Centre, Jagiellonian University, PL--30--059 Cracow, Poland}
\maketitle


One of the key problems in studying complex systems is answering the question of stability. The standard linearization approach relies heavily on the large time asymptotics and can be misleading for intermediate times. This is especially pronounced when systems develop transient behaviour -- the analysis based on eigenvalues loses its significance and different approach is needed. This shortcoming in physics literature can be traced back to the work of Orr \cite{Orr1907:NONNORMFLUIDS} in the hydrodynamical context. Since then, similar ideas were revived in the context of fluid dynamics \cite{TTRD1993:HYDRONONNORM,Sch2007:NONMODALSTAB}, plasma physics \cite{CBP2009:TRANSPLASMA,RS2014:NONMODALPLASMA}, diffusion in porous media \cite{RCPSZ2008:NONMODALPOROUS} or pattern formation \cite{RCDL2011:TRANSIENTPATTERNS,BJG2017:NONORMPATTERN}. Further motivation for this work is rooted in the ecological literature on biological networks \cite{May1972:ECOLOGY,NC1997:TRANSECOL,TA2015:STABCOMPL40}.

The physical mechanism of transient behaviour is both relatively simple and quite general. It needs the system's components to interact asymmetrically and be stabilized by an effective dissipation mechanism. Asymmetry is indispensable since only then inter-eigenmodes fluxes of "energy" can be formed. Such an unbalanced flow renders particular eigenmodes overpopulated or amplified. Crucially, this mechanism does not break the overall stability -- the dissipation eventually wins over and the amplification effect is only temporary or \textit{transient}.

In the paper we focus on such transient trajectories for a system of non-linear ODE's. On an elementary example we show how eigenvalue-based linearization technique becomes misleading and how simultaneously the transient property is developed. We utilize the May-Wigner model to include this mechanism and inspect its generic features. We find a new transient regime where transient trajectories are present although uncommon. Based on these findings, we identify relevant degrees of freedom driving the transient behaviour and propose a natural extension of the May-Wigner model. Such a modification leads to a generic transient behaviour arising as a robust transient phase.
\begin{center}
\begin{figure}[h]
\includegraphics[scale=.4]{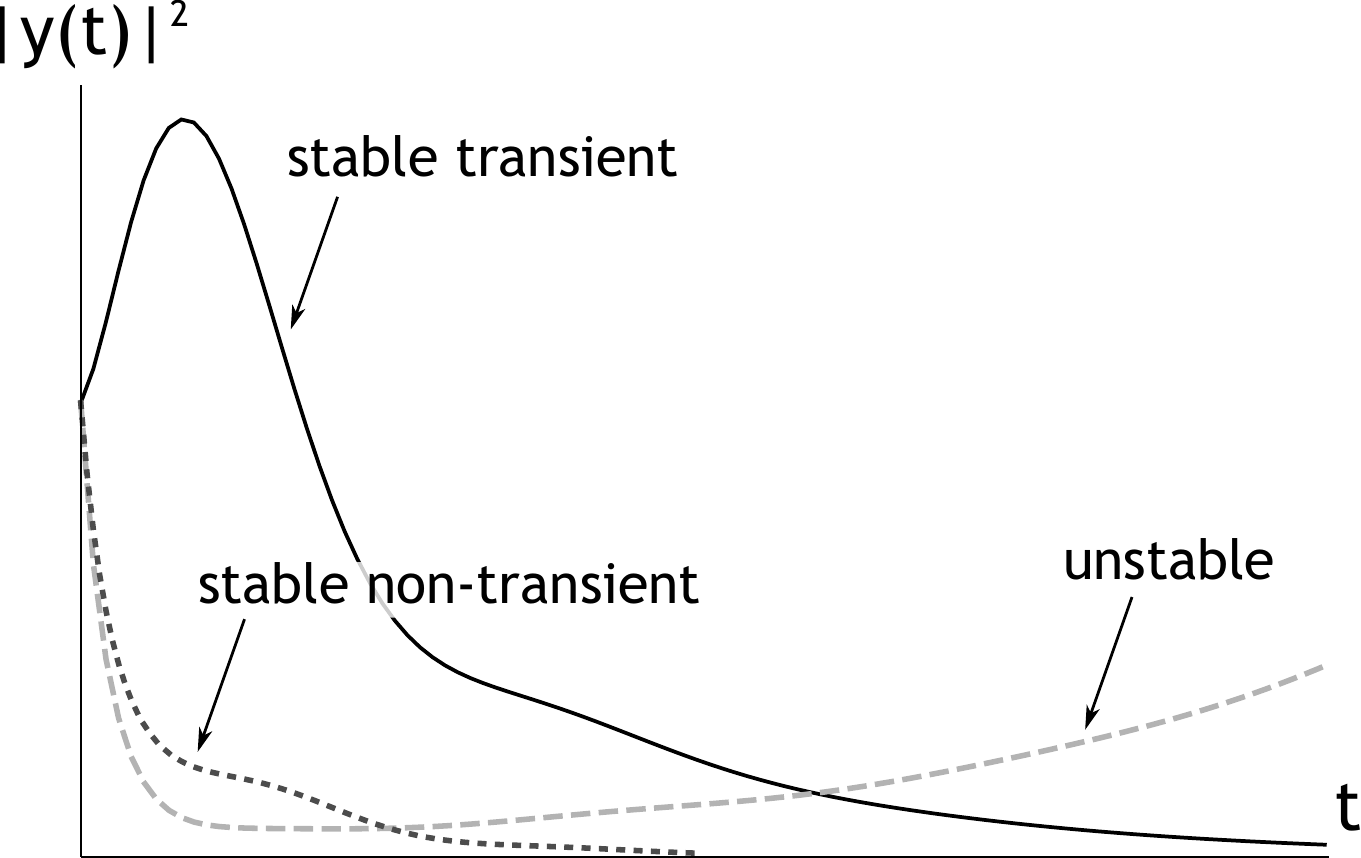}
\caption{The norms $|y(t)|^2 = \sum_i |y_i(t)|^2$ of all solutions to Eq. \eqref{eqlin} are divided into three types: stable transient, stable non-transient and unstable each marked by a solid-black, dotted-gray and dashed-gray lines respectively. Both stable classes reach zero asymptotically as $t\to \infty$ but differ at intermediate times.  Transient class develops an amplification beyond the initial value $|y(0)|^2$ whereas a non-transient trajectory does not posses such characteristic. The unstable solutions grow as $t\to \infty$ and are defined by this feature alone.}
\label{fig1}
\end{figure}
\end{center}
\subsection{Stability of systems}
We focus on a typical complex dynamical system of a set of $N$ first-order non-linear ordinary differential equations:
\begin{align}
\label{eqmain}
\frac{dx_i}{dt} = f_i(x_1...x_N), \qquad i = 1...N,
\end{align}
where $x_i$ are the relevant degrees of freedom (neurons, concentrations of chemical compounds, species, etc.) and the non-linear functions $f_i$ encode the interactions (f.e. the Lotka-Volterra competitive predator-prey model for $f_i = x_i(1-\sum_{j} \alpha_{ij} x_j)$). 

\begin{center}
\begin{figure}[h]
\includegraphics[scale=.35]{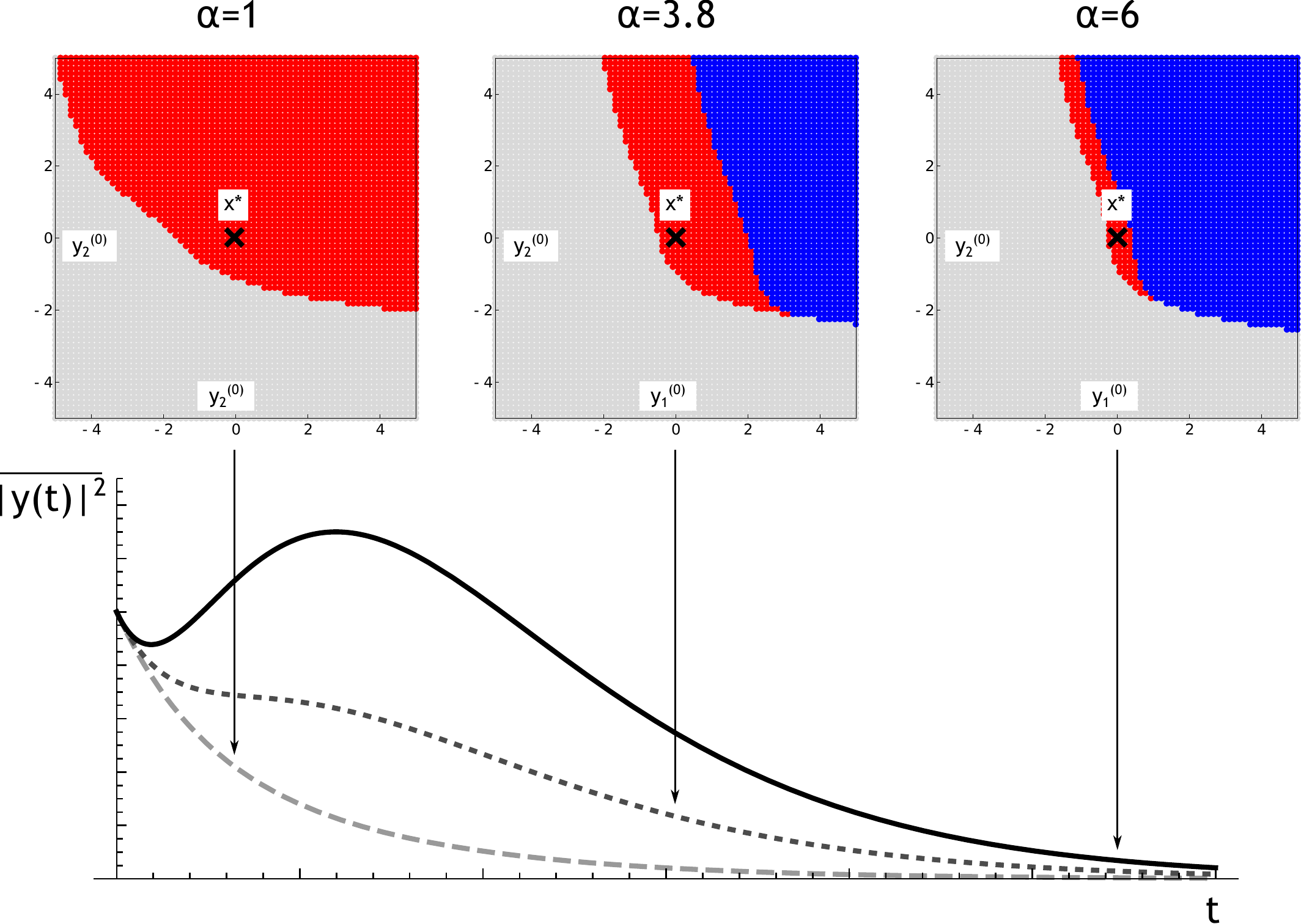}
\caption{\textbf{Top:} A numerical study of the stability of a system defined by Eq. \eqref{eqexample}. The solutions for different values of initial conditions $y_0^{(1)},y_0^{(2)}$ are plotted on the axes, inspected and sorted into three regimes -- an unstable basin and regions of stability of $x^*=(0,0)$ and $x^{**}$. They are colored by gray, red and blue points respectively, a black cross marks the stable point $x^*$ around which we conduct the linearization. As we increase $\alpha$, the $x^*$ stability basin shrinks considerably. By the H-G theorem, said point is stable for any positive $\alpha$ however in the full non-linear picture it becomes sub-dominant in comparison to the other regions showing how the locality assumption is a limiting factor and one need additional tools.
 \textbf{Bottom:} The linearized and averaged over $y_0$ solution to Eq. \eqref{eqlinexample} is equal to $\left < |y(t)|^2 \right >_{y_0} = e^{-3t} (\cosh t + \alpha^2 (\cosh t - 1))$ and plotted for three different values of $\alpha$. A norm first develops a bump for $t>0$ breaking the monotonicity which afterwards becomes a transient amplification. The cases for $\alpha=1$, $\alpha=3.8$ and $\alpha=6$ are depicted by dashed gray, dotted gray and solid black lines respectively. This change in the behaviour of the linearized solutions happens on par with the shrinkage of the $x^*$ stability basin depicted on the top plots.}
\label{fig2}
\end{figure}
\end{center}

If the form of functions $f_i$ is known, a question of stability is answered by a standard argument revised here briefly. In present analysis we ignore chaotic attractors or limit cycles and restrict to a simple binary notion of stability, the latter was addressed recently in \cite{IS2016:MAYWIGNER}. As a first step, we find all the points $f_i(x^*)=0$ at which the solutions remain constant in time. Next, we expand Eq. \eqref{eqmain} around a certain point $x^*$ from that set:
\begin{align}
x_i = x_i^* + y_i,
\end{align}
and find a linearized system of equations
\begin{align}
\label{eqlin}
\frac{d}{dt} y_i(t) = \sum_{j=1}^N M_{ij} y_{j}(t),
\end{align}
where $M_{ij} = \partial_{x_j} f_i(x)_{|x=x^*}$. According to the Hartman-Grobman theorem (H-G theorem) \cite{Tes2012:ODEDYNSYSTEMS}, the chosen point $x^*$ is stable if the real parts of the eigenvalues of $M$ are all strictly negative and unstable otherwise. The main assumption in the H-G theorem is that of locality -- the perturbation $y$ around a stable point should be small. We address its importance in an example considered in the following and presented in Fig. \ref{fig2}. 

To proceed, we define the norm of the solution $y_i$ of Eq. \eqref{eqlin} as $|y(t)|^2 = \sum_{i} |y_i(t)|^2$ and group them into into three groups:
\begin{enumerate}
\item \textit{stable non-transient} (or non-transient) when $|y(t)|^2 \overset{t \to \infty}{\longrightarrow} 0$ and $\underset{t}{\max} |y(t)|^2 = |y(0)|^2$, 
\item  \textit{stable transient} (or transient) when $|y(t)|^2 \overset{t \to \infty}{\longrightarrow} 0$ and $\underset{t}{\max} |y(t)|^2 \neq |y(0)|^2$,
\item \textit{unstable} when $|y(t)|^2 \overset{t \to \infty}{\longrightarrow} \infty$.
\end{enumerate}
Instances of these types are shown in Fig. \ref{fig1}. 

We present an example demonstrating the importance of locality assumption and simultaneously motivating this study. We define an $N=2$ dimensional non-linear system:
\begin{align}
\begin{cases}
\dot{x_1} = - x_1 + x_2^3 \\
\dot{x_2} = \alpha x_1 - 2 x_2 - x_1 x_2 - x_2^4
\end{cases},
\label{eqexample}
\end{align}
which has two relevant stable points: $x^* = (x_1^*,x_2^*) = (0,0)$ and $x^{**}$ (given implicitly). We linearize the system around $x^*$ and find the matrix:
\begin{align}
M_{|x^*} = \left ( \begin{matrix} - 1 & 0 \\ \alpha & -2 \end{matrix} \right ),
\end{align}
so that Eq. \eqref{eqlin} reads
\begin{align}
\begin{cases}
\dot{y_1} & = - y_1 \\
\dot{y_2} & = \alpha y_1 - 2y_2
\end{cases}.
\label{eqlinexample}
\end{align}
The resulting matrix $M_{|x^*}$ is in a triangular form, eigenvalues $-1,-2$ are strictly negative, the point $x^*$ is stable and so is the full system given by Eq. \eqref{eqexample}. We can solve the Eq. \eqref{eqlinexample} explicitly and find the norm of its solution $|y(t)|^2 = \left (y^{(0)}_2 -\alpha y^{(0)}_1 \right )^2 e^{-4t} + 2 y^{(0)}_1 \alpha e^{-3t} \left (y^{(0)}_2 - \alpha y^{(0)}_1 \right ) + e^{-2t} \left (y^{(0)}_1 \right )^2(1+\alpha^2 )$ depending on the initial value vector $y^{(0)} = \left (y^{(0)}_1,y^{(0)}_2 \right )$. From this formula one readily computes that for $\alpha > \frac{(y^{(0)}_1)^2 + 2(y^{(0)}_2)^2}{2 y^{(0)}_1 y^{(0)}_2}$, the transient behaviour of the norm $|y(t)|^2$ is present and absent otherwise. 

In Fig. \ref{fig2} we inspect the trajectories of both full and linearized system given by Eqs \eqref{eqexample} and \eqref{eqlinexample} as we vary the $\alpha$ parameter. We put emphasis on two features emerging in a correlated fashion -- shrinkage of the $x^*$ related basin of attraction of the full system (top plots of Fig. \ref{fig2}) and simultaneous development of transient dynamics in the linearized system (bottom plot of Fig. \ref{fig2}). In the spirit of previous studies on the linearized dynamics \cite{TTRD1993:HYDRONONNORM,Sch2007:NONMODALSTAB}, we hypothesize that $\alpha$ parameter (representing all of the non-eigenvalue degrees of freedom for $N>2$) drives both mechanisms. Therefore, extracting transient behaviour becomes particularly important when either the non-linear solution or the structure of the phase space is not known and so only the linearized information is accessible. Then, transient dynamics can be seen as a herald of (non-linear) instability present already in the linear regime. This observation is closely related to the study of basins of attraction by the so-called Lyapunov functions.

\subsubsection{Randomness and complexity}

Our aim is to study statistical features of transient phenomena highlighting its average features. To this end we chose a framework of random matrices as a unique insight into generic behaviour of such systems and is often treated as a first approximation or null-model analogous to a Gaussian distribution in univariate statistical analysis.

A matrix $M$ of size $N\times N$ introduced in Eq. \eqref{eqlin} is taken to be
\begin{align}
\label{mdef}
M = - \mu + X,
\end{align}
where $\mu$ is understood as a diagonal matrix with entries equal to $\mu>0$. It is used since the linearization procedure is computed at a stable point by assumption. In the following we consider both real and complex matrices $X$ denoted by an index $\beta=1$ and $\beta=2$ respectively. The matrix $X$ is random and drawn from a joint pdf:
\begin{align}
\label{pdf}
P_\beta(X) [dX]_\beta = c_\beta \exp \left ( -\frac{\beta N}{2\sigma^2} \text{Tr} X^\dagger X \right ) [dX]_\beta,
\end{align}
where $\sigma^2$ is the variance, the real matrix is decomposed as $X_{kl} = x_{kl}$ whereas the complex matrix reads $X_{kl} = x_{kl} + i y_{kl}$. The joint measure for the real case reads $[dX]_{\beta=1} \equiv \prod_{i,j=1}^N d x_{ij}$, for the complex case is $[dX]_{\beta=2} \equiv \prod_{i,j=1}^N d x_{ij} y_{ij}$ and the normalization constant $c_\beta^{-1} = \int P_\beta(X) [dX]_\beta $. A notation for the average over $X$ is given by
\begin{align}
\overline{O} = \left < O(X) \right >_X = \int [dX]_\beta P_\beta(X) O(X).
\end{align}

Lastly, we describe indicators of transient trajectories $y(t)$ as introduced in \cite{NC1997:TRANSECOL}. By inspecting definitions of trajectories depicted in Fig. \ref{fig1} one readily finds a good description of transient behaviour as the maximal possible amplification of the norm
\begin{align}
\label{amplifdef}
A =  \max_{t\geq 0} \frac{|y(t)|^2}{|y(0)|^2},
\end{align}
which identifies trajectory as stable non-transient if $A=1$, stable transient if $1<A<\infty$ and unstable if $A\to\infty$. Since for arbitrary systems it is hard to compute explicitly, instead a reactivity parameter $R$ was proposed:
\begin{align}
\label{reactivitydef}
R = \frac{1}{|y(0)|^2} \lim_{t\to 0} \frac{d|y(t)|^2}{dt}, 
\end{align}
as a measure of the \textit{initial} response of the system. By restricting to stable trajectories, we define transient behaviour if $R>0$ and lack thereof if $R<0$. The reactivity is an imperfect indicator -- truly transient trajectories can be misidentified as a non-transient. Since the opposite cannot occur, it systematically over-counts non-transient trajectories and numerical results suggest it is a small effect. 

We compute reactivity by writing down a formal solution to Eq. \eqref{eqlin}:
\begin{align}
\label{formsol}
\ket{y(t)} = e^{M t} \ket{y_0},
\end{align}
where we introduce an identification between the vectors $y_i$ and kets $(\ket{y})_i$ and denote the initial vector as $\ket{y_0} = \ket{y(0)}$. We plug Eq. \eqref{formsol} into Eq. \eqref{reactivitydef} and find:
\begin{align}
\label{Rformula}
R = \frac{\left < y_0 | (M^\dagger + M ) | y_0 \right >}{\left < y_0 | y_0 \right >},
\end{align}
where the braket notation dictates that $\left < y_0 | y_0 \right > = |y(0)|^2$. The two measures are related as the reactivity $R$ is a linear term in the expansion of the amplification $A$ around $t=0$, $A = \max_t \left ( 1 + R t + \mathcal{O}(t^2) \right )$. It therefore takes into account only the initial amplification.

Lastly, we address the treatment of initial conditions $\ket{y_0}$. A priori, we consider two scenarios -- an \textit{extreme case} where we chose a particular vector $y_0$ to maximize the quantity in question (f.e. reactivity) or a \textit{typical case} where we average over all initial conditions. We introduce a notation to designate both scenarios:
\begin{align}
\label{initaverages}
O_{\text{av}} & = \left < O(y_0) \right >_{y_0} = \int [d y_0]_\beta p_0(y_0) O(y_0), \\
O_{\max} & = \max_{y_0} O, 
\end{align}
with a flat measure $[dy_0]_{\beta}$ over real $\beta=1$ or complex $\beta=2$ initial vectors and $p_0$ denoting a prescribed pdf for the initial conditions. 

\subsection{Transient phase in the May-Wigner model}

The model defined by Eqs \eqref{mdef} and \eqref{pdf} was introduced in the seminal work of May \cite{May1972:ECOLOGY} to answer a key question in biological systems about the interplay between stability and complexity. The main finding is that there is an inherent (linear) instability of the system as we increase the complexity (matrix size $N$). Firstly, we review briefly this classic result and show how to extend the model to also include transient dynamics.

\begin{center}
\begin{figure}[h]
\includegraphics[scale=.5]{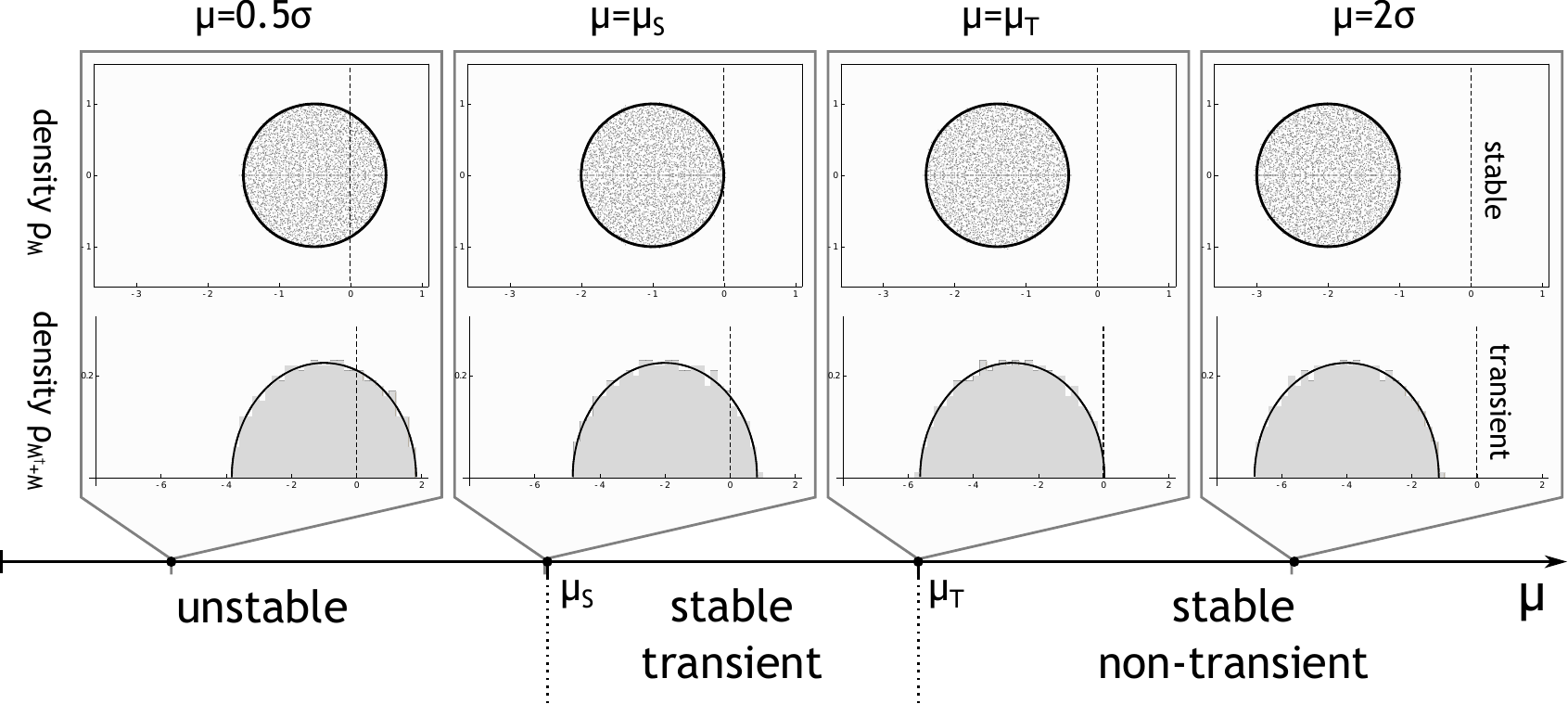}
\caption{Stable/unstable and transient/non-transient transitions are depicted as a function of the parameter $\mu$. The stability boundary is probed by the spectral density $\rho_M$ given by Eq. \eqref{girko} and plotted in the top row. The transient transition is probed with the density $\rho_{M^\dagger + M}$ and plotted in the bottom row. Plots of spectral densities and numerical simulations were drawn in solid lines and gray points/histograms. The boundaries for either transition are shown as dashed vertical lines. All three regimes are identified by the corresponding densities passing through the vertical boundaries. Simulations were conducted for real $\beta=1$ matrices of size $N=500$ and $\sigma=1$.}
\label{fig3}
\end{figure}
\end{center}

To recreate classic stability regimes we inspect the eigenvalue spectrum of $M$ as the matrix size grows to infinity $N\to \infty$. The asymptotic spectral density $\rho_{M}(x,y) = \lim\limits_{N\to \infty} \frac{1}{N} \left <\text{Tr} \delta (x+iy - M) \right >_X$ is given by the circular law \cite{Gir1985:CIRCULARLAW}:
\begin{align}
\label{girko}
\rho_{M}(x,y)  = \frac{1}{\pi \sigma^2} \theta \left (\sigma^2 - (x+\mu)^2-y^2 \right ),
\end{align}
where $\theta$ is the Heaviside theta function confining the eigenvalues inside a circle of radius $\sigma$ centered around $(-\mu,0)$. In the large $N$ limit, the result \eqref{girko} is valid for both values of $\beta=1,2$. The standard stability criterion based on the H-G theorem means that all eigenvalues of $M$ have real parts less than zero. In geometric terms, we keep the circular support of $\rho_M$ from crossing the $x=0$ line to stay in the stable regime. The stability/instability transition thus occurs along with the crossing and there are two equivalent ways of achieving that -- by increasing the radius $\sigma$ or by moving the center point $\mu$. In further discussion we focus on the latter formulation and restrict to modifying the $\mu$ parameter. Thus, we identify two regimes -- stable for $\mu>\mu_S$ and unstable for $\mu<\mu_S$ with $\mu_S = \sigma$ and depict this transition in Fig. \ref{fig3}. 

By inspecting transient character of trajectories, stable regime is additionally split into transient and non-transient parts. A boundary is defined by the maximal reactivity of Eq. \eqref{reactivitydef} with $\left < R_{\max} \right >_X > 0$ for stable transient and $\left < R_{\max} \right >_X < 0$ for stable non-transient regime. A similar boundary was studied in \cite{TA2014:TRANSECOL} and in our context it is an example of an extreme scenario in the sense of Eq. \eqref{initaverages}. As reactivity of Eq. \eqref{reactivitydef} is a Rayleigh quotient, it can be shown that $R_{\max}$ is given by the largest eigenvalue of $M^\dagger + M$:
\begin{align*}
R_{\max} = \max_{y_0} \frac{\left < y_0 | (M^\dagger + M ) | y_0 \right >}{\left < y_0 | y_0 \right >} = \lambda_{\max} \left ( M^\dagger + M \right ).
\end{align*}
Because the matrix $M^\dagger + M$ is symmetric for $\beta=1$ (or Hermitian for $\beta=2$), its spectrum in the large $N$ limit is a translated Wigner's semicircle $\rho_{M^\dagger + M}(\lambda) = \frac{1}{4\pi\sigma^2} \sqrt{8\sigma^2 - (\lambda+2\mu)^2}$. Using this result we read out the rightmost edge and so the averaged maximal reactivity in the large $N$ limit reads
\begin{align}
\label{r1max}
\lim_{N\to \infty} \left < R_{\max} \right >_X = -2\mu + 2 \mu_T,
\end{align}
where $\mu_T = \sqrt{2} \sigma$. We identify the stable transient regime for $\mu < \mu_T$ and stable non-transient regime for $\mu > \mu_T$ and presented it in Fig. \ref{fig3}. 

Although we understand stable and unstable regimes quite well, it remains to inspect further the novel transient regime when $\mu \in (\mu_S,\mu_T)$. A natural question to ask is how \textit{abundant} transient amplification is in the ensemble of trajectories. It is relevant since, by using a different criterion based instead on a reactivity averaged over the initial conditions $R_\text{av}$ given by Eq. \eqref{initaverages}, we find $\left < R_{\text{av}} \right >_X = -2\mu$. As $\mu>0$, $\left < R_{\text{av}} \right >_X$ is always negative and does not predict a transient behaviour. 

\subsubsection{Density and abundance of transient trajectories} 

To inspect the question of abundance of transient trajectories, we define a probability density for the reactivity $g(r) = \delta (r-R(y_0))$ and consider both maximal and typical densities:
\begin{align}
\overline{g_{\max}} (r) = \left < \max_{y_0} g(r) \right >_X, \quad \overline{g_{\text{av}}}(r) = \left < g(r) \right >_{X,y_0}.
\end{align}
Although the averaging over $X$ and $y_0$ is interchangeable $\overline{g_{\text{av}}} = \overline{g}_{\text{av}}$, the $\max$ operation and average is not $\overline{g_{\max}} \neq \overline{g}_{\max}$. Firstly we compute the density $\overline{g}$ averaged only over $X$:
\begin{align}
\label{gavint}
\overline{g} = c_\beta \int [dX]_\beta P_\beta(X) \delta (r-R(y_0)),
\end{align}
where an implicit dependence of $\overline{g}$ on $\beta$ is assumed. We use the delta function representation $\delta(x) = (2\pi)^{-1} \int dp e^{ipx}$, rewrite $ip(r-R) = ip(r+2\mu) - i\alpha \text{Tr} Y (X+X^\dagger)$ where $\alpha = p (\text{Tr} Y^2)^{-1}$ and set $Y = \ket{y_0} \bra{y_0}$. We compute the integral \eqref{gavint} by completing the square:
$ i \alpha \text{Tr} Y (X+X^\dagger) + \frac{\beta N}{2\sigma^2} \text{Tr} X^\dagger X = \frac{\beta N}{2\sigma^2} \text{Tr} \left (X^\dagger + \frac{2i\sigma^2 \alpha}{\beta N} Y \right )\left (X + \frac{2i\sigma^2 \alpha}{\beta N} Y \right ) +\frac{2\alpha^2\sigma^2}{\beta N} \text{Tr} Y^2 $. The result is a quadratic Fourier integral:
\begin{align}
\overline{g}(r) = \frac{1}{2\pi} \int dp e^{ip(r+2\mu)} e^{-\frac{2\sigma^2 p^2}{N} },
\end{align}
which no longer depends on the initial values $Y$ as $\text{Tr} Y^2 = (\text{Tr} Y)^2$. The remaining integration gives a Gaussian distribution
\begin{align}
\label{gav}
\overline{g}(r) = \frac{1}{\sqrt{2\pi \sigma_{\beta,R}^2}} e^{-\frac{(r+2\mu)^2}{2 \sigma_{\beta,R}^2}},
\end{align}
with mean $-2\mu$ and variance $\sigma_{\beta,R}^2 = \frac{4\sigma^2}{\beta N}$. Eq. \eqref{gav} is already the typical reactivity density $\overline{g_{\text{av}}}$ as it does not depend on the choice of initial conditions $y_0$, and so trivially $\overline{g} = \overline{g_{\text{av}}}$.

We turn to the extreme reactivity density which is related to the largest eigenvalue of $X^\dagger + X$:
\begin{align*}
\overline{g_{\max}}(r) & = \left < \delta \left (r+2\mu - \lambda_{\max}(X^\dagger + X) \right ) \right >_X \\
& = \frac{d}{dr} \left < \theta \left ( \mu + \frac{r}{2} - \lambda_{\max} \left ( \frac{X^T + X}{2} \right ) \right ) \right >_X,
\end{align*}
where an implicit dependence of $\overline{g_{\max}}$ on $\beta$ is assumed. In the literature on extreme value statistics \cite{TW2009:TWINTRO}, one defines a cumulative distribution function $F_{N,\beta}(t) = \int [dH] e^{-\frac{\beta}{2} \text{Tr} H^2}\theta \left ( t - \lambda_{\max}(H) \right )$ of the largest eigenvalue. By a simple rescaling, the extreme reactivity density therefore reads
\begin{align}
\label{gmax}
\overline{g_{\max}}(r) = \frac{d}{dr} F_{N,\beta}\left (\frac{\sqrt{N}}{\sigma}\left (\mu + \frac{r}{2} \right ) \right ).
\end{align}

In this case, the order of operations is crucial -- taking first the average over $X$ and then maximizing will reduce to the average scheme as $\overline{g}_{\max} = \overline{g}$. This is expected as the extreme scenario of any observable $O$ can be realized as an average scheme given by Eq. \eqref{initaverages} but over a particular point-source density $\rho_0 \sim \delta (y_0 - v_{\max}(X))$ dependent on the maximal eigenvector $v_{\max}$ of $X$ itself. If so, the two averages no longer commute.
\begin{center}
\begin{figure}[h]
\includegraphics[scale=.5]{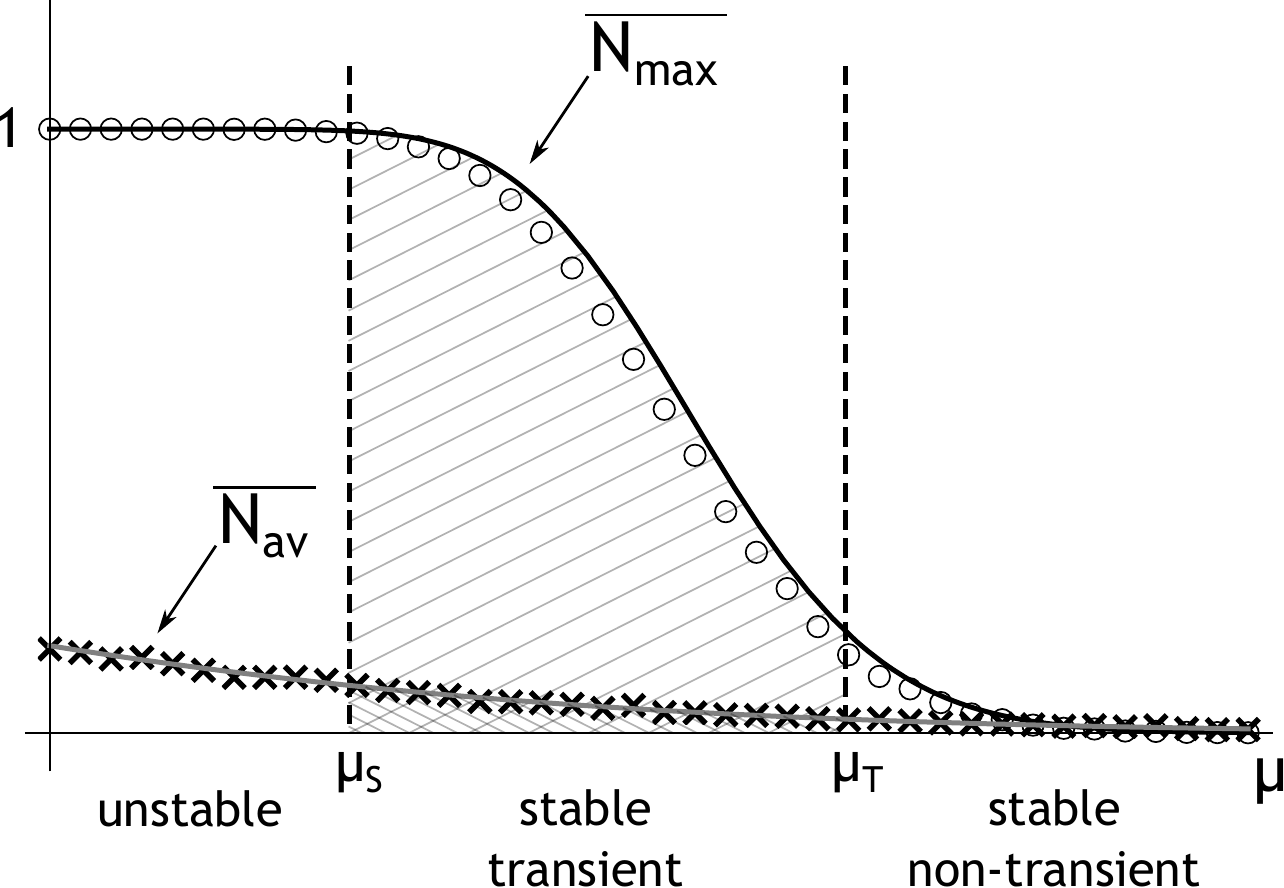}
\caption{Numerical and analytical plots of transient trajectories abundances $\overline{N_{\text{av}}}$ and $\overline{N_{\max}}$ as a function of the stability parameter $\mu$. Boundaries $\mu_S = \sigma$ and $\mu_T = \sqrt{2}\sigma$ delineate between unstable, stable transient and stable non-transient regimes. Numerical results are shown for $\overline{N_{\max}}$ as circles ($N=20$) and for $\overline{N_{\text{av}}}$ as crosses ($N=2$). Analytical results of \eqref{navlarge} and \eqref{nmax} are plotted as solid gray and black lines.}
\label{fig4}
\end{figure}
\end{center}
The abundances of transient trajectories are found as tail distributions of previously computed densities given in Eqs \eqref{gav} and \eqref{gmax}:
\begin{align}
\overline{N_{\max}} = \int_0^\infty \overline{g_{\max}}(r) dr, \quad \overline{N_{\text{av}}} = \int_0^\infty \overline{g_{\text{av}}}(r) dr,
\end{align}
which results in an implicit formula for the $\overline{N_{\max}}$ and an explicit one for $\overline{N_{\text{av}}}$:
\begin{align}
\overline{N_{\max}}(\mu) & = 1 - F_{N,\beta}\left (\frac{\sqrt{N}\mu}{\sigma} \right ), \\
\overline{N_{\text{av}}}(\mu) & = \frac{1}{2} \text{erfc} \left ( \sqrt{ \beta N} \frac{\mu}{\mu_T} \right ).
\end{align}
We compute the asymptotic forms of both abundances as $N\to \infty$. The abundance of a typical transient trajectory is asymptotically Gaussian:
\begin{align}
\label{navlarge}
\overline{N_{\text{av}}} \sim e^{-\beta N\left (  \frac{\mu}{\mu_T}\right )^2 }.
\end{align}

Although the abundance of an extreme transient trajectory is expressed in terms of the $F_{N,\beta}$ function which is not known in an explicit form, instead we cite two results valid in the $N\to \infty$ limit. To this end, we set $\mu = \mu_T - \Delta$ and inspect deviations around a typical value $\mu_T$ on different scales $\Delta$. The asymptotic formula of $\overline{N_{\max}}$ for large deviations $\Delta \sim \mathcal{O}(1)$ was found in \cite{DM2006:EXTREMEGBE1,DM2008:EXTREMEGBE2} as:
\begin{align}
\label{largedev}
F_{N,\beta}\left (\sqrt{2N} - \frac{\sqrt{N}\Delta}{\sigma} \right ) \sim e^{-\beta N^2 \phi(\Delta/\sigma-\sqrt{2})}, 
\end{align}
with $\phi(x) = \frac{1}{108} \left \{ -x^4+36 x^2+\sqrt{x^2+6} \left(x^3+15 x\right) + \right .$  $\left .+ 27 \left(\log (18)-2 \log \left(\sqrt{x^2+6}-x\right)\right)\right \}$. We note the same formulas arose in a discussion of the \textit{symmetric} May-Wigner model near its stability transition in \cite{MS2014:EXTREMEEVS} which is however not equivalent to our case. For small perturbations $\Delta \sim \mathcal{O}(N^{-2/3})$ we cite another result:
\begin{align}
\label{smalldev}
F_{N,\beta}\left (\sqrt{2N} - \frac{\sqrt{N}\Delta}{\sigma} \right ) \sim F_{\beta} \left (-N^{\frac{2}{3}} \frac{\sqrt{2}\Delta}{\sigma} \right ), 
\end{align}
where $F_{\beta}$ is the Tracy-Widom distribution \cite{TW1993:TW1,TW1994:TW2}. The abundance of extreme transient trajectories is therefore given as:
\begin{align}
\overline{N_{\max}}(\mu_T-\Delta) \sim 1 - \left \{ 
\begin{tabular}{ll}
$e^{-\beta N^2 \phi(\Delta/\sigma-\sqrt{2})}$ &, $\Delta \sim \mathcal{O}(1)$ \\ 
$F_{\beta} \left (-N^{\frac{2}{3}} \frac{\sqrt{2}\Delta}{\sigma} \right )$ &, $\Delta \sim \mathcal{O}(N^{-\frac{2}{3}})$
\end{tabular} \right . ,
\label{nmax}
\end{align}

We plot both $\overline{N_{\text{av}}}$ and $\overline{N_{\max}}$ given by Eqs \eqref{navlarge} and \eqref{nmax} in Fig. \ref{fig4} along with numerical results. The abundance of extreme transients $\overline{N_{\max}}$ is directly related to the transient/non-transient boundary at $\mu=\mu_T$ as it becomes a sharp theta function $\overline{N_{\max}} \sim \theta (\mu_T-\mu)$ in the (global) $N\to \infty$ limit. For large and intermediate values of $N$, the maximal abundance $\overline{N_{\max}}$ increases rapidly as we traverse the $\mu_T$ boundary and reaches unity upon entering the unstable regime near $\mu_S$. The nature of the abundance of typical transients $\overline{N_{\text{av}}}$ is different -- it varies between $1/2$ for $\mu\to 0$ and $0$ if $\mu\to \infty$ however reaches zero in the transient regime between $\mu_S$ and $\mu_T$ quite rapidly. Moreover, as the size of the matrix $N$ increases, $\overline{N_{\text{av}}}$ approaches zero for all values of $\mu$ according to Eq. \eqref{navlarge} and does not result in a transition. Additionally, for any finite $N$ we find the average abundance being considerably smaller than the extreme one. 

We recapitulate these two complementary viewpoints: 1) transient regime for intermediate parameters $\mu \in (\mu_S, \mu_T)$ is found as the abundance of extreme trajectories $\overline{N_{\max}}$ is close to unity. Moreover, it increases with the system size $N$ and reaches certainty for formally infinite systems. 2) On the other hand, according to the abundance of average trajectories $\overline{N_{\text{av}}}$, the number of typical transient trajectories is relatively small when the initial conditions are not especially tailored. In fact, their number on average decreases rapidly with the growth of the systems' size as shown in Eq. \eqref{navlarge}.

Main conclusion is that although transient trajectories are (potentially) present in the whole transient regime $\mu \in (\mu_S, \mu_T)$ as shown by the behaviour of $\overline{N_{\max}}$, they are otherwise uncommon as dictated by $\overline{N_{\text{av}}}$. 
 
\subsubsection{Generators of transient behaviour}

Up to now we have considered the May-Wigner model of Eq. \eqref{eqlin} with matrices drawn from Eq. \eqref{pdf} and found that when $\mu \in (\mu_S,\mu_T)$, transient trajectories are present although rare. Drawn by the interest of the transient behaviour itself, we ask a related question -- although in May-Wigner model these trajectories are not found often, which features of a matrix we can tweak so that it produces generic transient trajectories. To put it differently, how to amplify the abundance $\overline{N_\text{av}}$ and simultaneously stay in the stable regime. To this end, we recall the definition of reactivity given in Eq. \eqref{reactivitydef}:
\begin{align}
\label{Ragain}
R = -2 \mu + \frac{\left < y_0 | (X^\dagger + X ) | y_0 \right >}{\left < y_0 | y_0 \right >},
\end{align}
where the presence or absence of transient behaviour was defined by the sign of $R$. If $X$ is drawn from Eq. \eqref{pdf}, we have shown previously that although $\left < R_{\max} \right >_X $ can be positive (which sets the scale $\mu_T$), on average $\left < R \right >_X = - 2\mu$ is always negative and thus no typical transient behaviour is present. To circumvent this we need to modify the matrix measure \eqref{pdf} accordingly. 

Our aim is to render the reactivity \eqref{Ragain} positive. For pure May-Wigner models defined by Eq. \eqref{pdf}, it is always negative since both $\left < X \right >_{X}$ and $\left < X^\dagger \right >_{X}$ are zero. A simplest route of introducing a pdf with a non-zero mean $\left < X \right >'_{X} = X_0 \neq 0$ does not produce satisfactory result since then also the eigenvalue density of $X$ is modified resulting in an instability. The way out is to freeze the eigenvalues and tweak the remaining degrees of freedom. To achieve that, we introduce a Schur decomposition \cite{HJ1990:LINEARALGEBRABOOK}:
\begin{align}
\label{schur}
X = O (Z+T)O^\dagger,
\end{align}
where the matrix $O$ is orthogonal ($\beta=1$) or unitary ($\beta=2$), $Z$ is a diagonal matrix with eigenvalues and $T$ is a strictly upper-triangular matrix encoding the non-orthogonality of the eigenbasis, hereafter we will refer to it as the non-orthogonality matrix. 

For $\beta=1$, both $Z$ and $T$ have a block structure determined by the character of eigenvalues which are either purely real or form complex conjugate pairs. If we assume there are $k$ purely real eigenvalues and $\frac{N-k}{2}$ conjugate pairs of complex eigenvalues, the blocks of $Z$ and $T$ have dimensions $k\times k$ and $\frac{N-k}{2}\times \frac{N-k}{2}$ on the diagonal and $k\times \frac{N-k}{2}$ on the off-diagonal. The individual entries are either numbers or $2\times 2$ matrices on the diagonal and $2\times 1$ vectors on the off-diagonal. 

For $\beta=2$, both $Z$ and $T$ have a simple structure - $Z$ is diagonal filled with eigenvalues and $T$ is strictly upper-triangular.

\begin{center}
\begin{figure}[h]
\includegraphics[scale=.9]{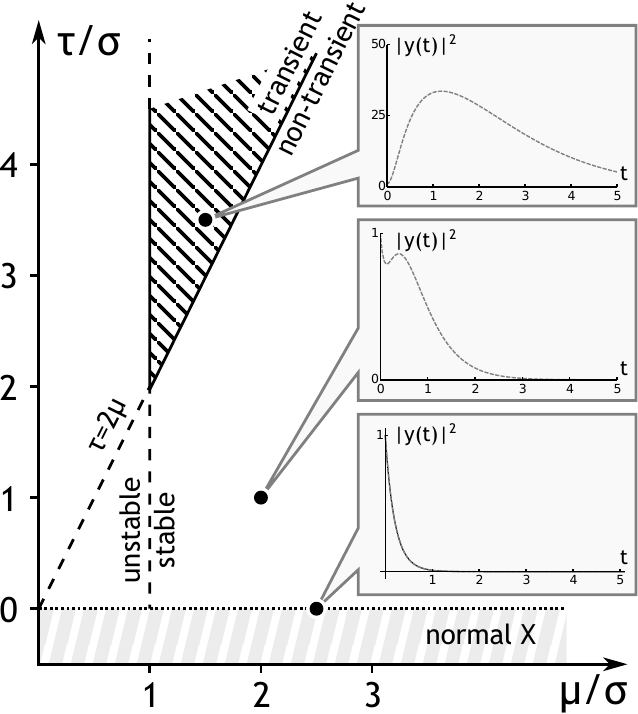}
\caption{A phase diagram of the fixed $T_0$ model defined by Eq. \eqref{pdfFixedT} with parameter $\tau$ defined in Eq. \eqref{tau} and $\mu$ given in Eq. \eqref{mdef}. Straight lines $\mu=\sigma$ and $\tau = 2\mu$ mark stability and transient boundaries, the dashed intersection area is the transient stable regime and at $\tau=0$ we plot a horizontal dotted line denoting the normal matrix regime. From top to bottom, three inset plots were evaluated at phase points $(\mu,\tau)=(1.5,3.5)$, $(2,1)$ and $(2.5,0)$ and correspond to a transient stable, non-transient stable and normal trajectory respectively. The plots were obtained for $N=40$, $(T_0)_{ij} = \sigma\alpha ~\delta_{i1}\delta_{j2}$ for a fixed initial value $y_0$ and $\alpha$ being equal to $125, 29$ and $0$ respectively. The bottom plot admits an analytic form of Eq. \eqref{normavplug}.}
\label{fig5}
\end{figure}
\end{center}

In the $\beta=1$ case, this decomposition induces a change of variables in the measure \eqref{pdf} computed in \cite{LS1991:REALGINIBRE2,Ede1997:REALGINIBRE} and for $\beta=2$ it is trivial. Because both 1) the Jacobian of the transformation \eqref{schur} factorizes into $Z$ and $T$ dependent parts and 2) the Gaussian factor of Eq. \eqref{pdf}
\begin{align}
\text{Tr} X^\dagger X = \text{Tr} Z^\dagger Z + \text{Tr} T^\dagger T
\end{align}
factorizes for both $\beta=1,2$, the eigenvalue matrix $Z$ and the non-orthogonality matrix $T$ fully decouple and can vary independently. Using Eq. \eqref{schur}, one finds that the averaged reactivity 
\begin{align}
 \left < R \right >_X = -2\mu + \frac{1}{\left < y_0 | y_0 \right >} & \Big ( \left < y_0 | \left < Z^\dagger + Z \right >_X | y_0 \right > +  \nonumber \\
 & + \left < y_0 | \left < T^\dagger + T \right >_X | y_0 \right > \Big )
 \end{align}
is likewise decoupled. The first term $\left < Z^\dagger + Z \right >_X$ is zero when $X$ is drawn from Eq. \eqref{pdf} as can be shown by a symmetry argument -- for $\beta=1$ we reflect real eigenvalues $\lambda_i \to -\lambda_i$ along with the real parts of complex pairs $\text{Re} z_i\to -\text{Re} z_i$ and for $\beta=2$ we just set $z_i \to -z_i$. Although the second part $\left < T^\dagger + T \right >_X$ is also zero when averaged over Eq. \eqref{pdf}, it does not need to be the case. In particular, we can fix $T$ by a constraint $\delta (T- T_0)$ nad define a fixed $T_0$ May-Wigner model:
\begin{align}
\label{pdfFixedT}
\tilde{P}_\beta(X;T_0) [dX]_\beta = c'_\beta \delta (T - T_0) P_\beta(X) [dX]_\beta,
\end{align}
where $P_\beta(X)$ was introduced in Eq. \eqref{pdf} and $c'_\beta$ is the appropriate constant.  When averaged over Eq. \eqref{pdfFixedT} denoted as $\left < \cdots \right >_{\tilde{P}_\beta}$, we find an average reactivity
\begin{align}
\left < R \right >_{\tilde{P}_\beta} = - 2 \mu + \frac{\left < y_0 | T_0^\dagger + T_0 | y_0 \right >}{\left < y_0 | y_0 \right >}.
\end{align}
Due to independence of $Z$ and $T$, introducing a pdf in Eq. \eqref{pdfFixedT} does not change the spectrum of $X$. We define an external field/non-orthogonality parameter:
\begin{align}
\label{tau}
\tau = \frac{\left < y_0 | T_0^\dagger + T_0 | y_0 \right >}{\left < y_0 | y_0 \right >},
\end{align}
so that for a fixed $T_0$ model given by Eq. \eqref{pdfFixedT}, a generic transient behaviour is found when $\tau > 2 \mu$ and absent otherwise. The resulting phase diagram is shown in Fig. \ref{fig5}.

Now the parameter $\tau$ depends on the initial condition $y_0$. In particular, if the average $\left < \tau \right >_{y_0}$ is taken over the initial value vectors drawn from a symmetric density $p_0(y_0) = p_0(-y_0)$, it will vanish. A non-zero contribution is however produced if any asymmetry is present in $p_0$ or $y_0$ is held fixed. This was absent previously as the order parameter $\left < R \right >_X$ was completely decoupled from the initial condition $y_0$.
We interpret $T_0$ as a "field" conjugated to the order parameter $\left < R \right >_X$ akin to the the magnetic field and magnetization. Since the non-orthogonality matrix $T_0$ drives the transient phase transition, we consider the unperturbed system of $T_0=0$ in Eq. \eqref{pdfFixedT}. It describes a normal matrix model, defined also by the condition $[X^\dagger,X]=0$ and considered mostly when the matrix $X$ is complex \cite{CZ1998:NORMALCORR,WZ2003:NORMALMATRIX,TBA+2005:NORMALRMM}. In this particular case, we compute an exact formula for the average norm:
\begin{align*}
\left < |y(t)|^2 \right >_{\tilde{P}_\beta} = e^{-2\mu t} \int [dX]_\beta \tilde{P}_\beta(X;0) \text{Tr} \left ( Y e^{X^\dagger t}e^{X t} \right ).
\end{align*}
where we have used the formal solution given by Eq. \eqref{formsol} and $Y = \ket{y_0} \bra{y_0}$. Since $T_0=0$ we obtain $X = O Z O^\dagger$ by the Schur decomposition and find 
\begin{align}
 \left < |y(t)|^2 \right >_{\tilde{P}_\beta} & = e^{-2\mu t}  \int [dZ]_\beta P'_\beta(Z) \times \nonumber \\
 & \times \int [dO]_\beta \sum_{k,l,n}Y_{kl} O_{ln} e^{2\text{Re} z_n t} (O^\dagger)_{nk},
\end{align}
where $P'_\beta(Z)$ is the eigenvalue pdf for both values of $\beta=1,2$. The unitary/orthogonal integral is computed as $\int O_{ij} O^\dagger_{kl} [dO]_\beta = \frac{1}{N} \delta_{il} \delta_{jk}$ and the result reads:
\begin{align}
\label{normav}
 \left < |y(t)|^2 \right >_{\tilde{P}_\beta} = |y_0|^2 e^{-2\mu t} \int d^2 z \tilde{\rho}_{X}(z) e^{2t\text{Re} z},
\end{align}
where $\tilde{\rho}_{X}(z) = N^{-1} \left < \sum_{i=1}^N \delta^2 (z - z_i) \right >_{P'_\beta(Z)}$ is the spectral density of the normal matrix model. 

In the large $N$ limit, the spectral density of a normal matrix $\tilde{\rho}_{X}$ also forms a circular law given in Eq. \eqref{girko} with $\mu=0$. We plug it into Eq. \eqref{normav} and find the average norm in the large $N$ limit as
\begin{align}
\label{normavplug}
\lim_{N\to \infty} \left < |y(t)|^2 \right >_{\tilde{P}_\beta} = |y_0|^2 \frac{I_1(2t\sigma)e^{-2\mu t} }{t\sigma} .
\end{align}
It is monotonically decreasing (as shown in the bottom inset of Fig. \ref{fig5}) and does not present transient behaviour in accordance with previous results. 

\subsubsection{Non-orthogonality matrix and eigenvectors}

The non-orthogonality matrix $T$ present in the Schur decomposition given by Eq. \eqref{schur} is a crucial element in the development of transient behaviour. We will show in what sense the matrix $T$ is a measure of non-orthogonality and how it is related to other eigenvector related phenomena. In this section we restrict to the complex $\beta=2$ case. To this end, we diagonalize the matrix by a similarity transformation $S$:
\begin{align}
\label{similtrans}
X = S Z S^{-1},
\end{align}
where $S^{-1}_{ij}$ is composed of left eigenvectors  $\bra{L_i}_j$ and $S_{ji}$ of right eigenvectors $\ket{R_i}_j$:
\begin{align}
\bra{L_i} X = \bra{L_i} z_i, \quad X \ket{R_i} = z_i \ket{R_i}.
\end{align}
Left and right eigenvectors are bi-orthogonal $\braket{L_i | R_j} = \delta_{ij}$ but not orthogonal in each space separately i.e. $\braket{L_i | L_j} \neq \delta_{ij}$ and $\braket{R_i | R_j} \neq \delta_{ij}$. In terms of the matrix $S$, these two relations are rewritten as $\braket{L_i | L_j} = (S^\dagger S)_{ij}$ and $\braket{R_i | R_j} = (S^\dagger S)^{-1}_{ij}$. A formal relation between $S$, $T$ and $Z$ is found by juxtaposing Eqs \eqref{schur} and \eqref{similtrans}:
\begin{align}
O^\dagger S(Z+T) = Z O^\dagger S.
\end{align}
If $T=0$ and eigenvalues are non-degenerate we find $O^\dagger S \sim 1$, $S$ becomes an unitary matrix with left and right eigenvectors rendered orthogonal. For non-zero $T$, one finds recurrence relations between the orthogonality matrices $S^\dagger S$, $Z$ and $T$ \cite{CM1998:EIGENVECTORPRL}. These orthogonality matrices are used in the definition of a 1-point eigenvector correlation function
\begin{align}
\frac{1}{N} \left < \sum_{i=1}^N O_{ii} \delta(z-z_i) \right >_X
\end{align}
with weight $O_{ii} = (S^\dagger S)_{ii} (S^\dagger S)^{-1}_{ii}$. This object becomes a necessary ingredient in hydrodynamical description of complex dynamical matrices as established in \cite{BGN+2015:NHLONG,BGN+2014:NHSHORT}. The $O_{ii}$'s are also related to the eigenvalue condition number $\kappa(z_i)$ as shown in \cite{BNST2017:EVECS1}.

\subsection{Conclusions}

In this paper we characterize transient phenomena in generic complex systems. Motivated by both physics and interdisciplinary applications, we argue that transient behaviour is complementary to the stability analysis and hints at non-linear features already on the linear level. 

A seminal May-Wigner model is introduced as an example where we discuss transient dynamics. Based on the reactivity defined as an initial response of the system, we identify seminal unstable and stable regimes and find the latter to be additionally split into transient and non-transient regions. In the stable transient area, we compute the abundance (number) of transient trajectories in both extreme and typical choice of initial conditions. We conclude that although for certain special initial conditions trajectories do become transient, typically they do not show up. To amplify this typical transient behaviour, we introduce a modified May-Wigner model where only the non-orthogonality matrix $T_0$ is tweaked. This provides a model where a typical transient behaviour arises as a result of a non-zero value of $T_0$ with close relation to normal matrix models if $T_0$ vanishes. Lastly, the non-orthogonality matrix is discussed in relation to the eigenvector correlation function.

Transient dynamics have many faces -- they are often described as a destabilizing mechanism \cite{TTRD1993:HYDRONONNORM}. In particular, this study points toward identifying crucial characteristics important especially in the context of early-warning signals of transitions in complex systems \cite{Sch2009:EARLYCRIT}. In the studies of neural networks, it is responsible for memory effects \cite{GHS2008:MEMORY}. Paradoxically, in describing food-webs when the probed time span is relatively short wrt. the characteristic attenuation time of the transient, it can be reintrepreted as an effectively stable solution \cite{Has2004:TRANSECOL}. 

This papers sheds light on the relevant characteristics of transient behaviour, enabling the tools needed to assess the severity of transient behaviour in the system at hand and their ultimate stability. In the spirit of recent work \cite{FK2016:MAYSTAB}, in this work the phase diagram of May-Wigner models gets also refined to augment the stability questions. Additionally, transient features are robust when an additional structure is introduced into $X$ as presented in the \cite{TA2014:TRANSECOL}, pointing naturally into questions of universality. It is therefore an interesting point to study transient dynamics in general noise-plus-structure models of \cite{GG2016:EXACTSPECTR} with special emphasis on the application to neuronal networks \cite{AFM2015:NHCORRSOURCE,RA2006:NEURONRMT}.

\textbf{Acknowledgements.}

Author appreciates discussions with G. Schehr, S. Majumdar, M. A. Nowak and acknowledges the support of Grant DEC-2011/02/A/ST1/00119 of the National Centre of Science.

\bibliography{krkbib2015}{}

\end{document}